# Creation of User Friendly Datasets: Insights from a Case Study concerning Explanations of Loan Denials

Ajay Chander [1]   Ramya Srinivasan [1]

## Abstract

Most explainable AI (XAI) techniques are concerned with the design of algorithms to explain the AI's decision. However, the data that is used to train these algorithms may contain features that are often incomprehensible to an end-user even with the best XAI algorithms. Thus, the problem of explainability has to be addressed starting right from the data creation step. In this paper, we studied this problem considering the use-case of explaining loan denials to end-users as opposed to AI engineers or domain experts. Motivated by the lack of datasets that are representative of user-friendly explanations, we build the first-of-its-kind dataset that is representative of user-friendly explanations for loan denials. The paper shares some of the insights gained in curating the dataset. First, existing datasets seldom contain features that end users consider as acceptable in understanding a model's decision. Second, understanding of the explanation's context such as the human-in-the-loop seeking the explanation, and the purpose for which an explanation is sought, aids in the creation of user-friendly datasets. Thus, our dataset, which we call Xnet, also contains explanations that serve different purposes: those that educate the loan applicants, and help them take appropriate action towards a future approval. We hope this work will trigger the creation of new user friendly datasets, and serve as a guide for the curation of such datasets.

## 1. Introduction

Perhaps the analogy "data is the new oil" needs a re-examination. While it may appear that large amounts of data can solve a problem, bias in datasets coupled with missing, inaccurate or incomprehensible information can contribute to poor data quality. A recent study from Oxford Economics and ServiceNow showed that 51% of CIOs cite data quality as a substantial barrier to their company's adoption of AI (Service-Now, 2018). Even academic research (Tommasi et al., 2015; Torralba & Efros, 2017) has shown that data can significantly affect the performance of a machine learning system.

With growing concerns from customers and policy-makers about AI being a blackbox technology, the aforementioned problem gets even more aggravated. Consider, for example, an AI-based credit scoring system. In markets where credit risk scoring models are regulated and scrutinized, the AI models and credit decisions derived from them are strongly monitored and regulated to be explainable to lenders, regulators and consumers (FICO, 2018). While there have been several initiatives both from the government (Gunning, 2017) and a diverse set of industries (Kyndi, 2018; PWC, 2018) to make AI explainable, most state-of-the-art methods continue to provide explanations that mostly target the needs of AI engineers (Selvaraju et al., 2017).

Thus, there exists a gap between the explanations that are provided by AI models, and the explanations sought by various stakeholders. While there could be several factors that contribute to this gap, data is a significant contributor. In this paper, we explore strategies by which we might be able to bridge this gap. We discuss the process of curating datasets that serve the needs of stakeholders beyond AI scientists, and share the insights gained therein.

Motivated by the widespread adoption of AI in the Fintech industry, for the purposes of this study, we consider the use-case of explaining AI based loan decisions . From regulators, judges, and credit bureaus, to bank managers, investment stakeholders, and loan applicants, a variety of stakeholders may be in need of an explanation.

Depending on the role of the human-in-the-loop, the purpose an explanation should serve varies. For example, an AI scientist might want an explanation to understand what is the best debugging strategy if the model fails, or might want an explanation to know that the model is working as required, i.e. for validation purposes. Thus, an explanation such as "Person A was denied a loan because their profile was similar to person B who was also denied a loan" might be useful to the AI scientist in *validating* the model. But this kind of an explanation little helps a loan applicant!

A loan applicant, on the other hand, might want to know why they were denied the loan and perhaps what they can do to secure a future loan. Thus, the purpose of the explanation might be to *educate* the loan applicant or help them take appropriate *action* to secure a future loan.

Thus, some purposes of explanation could include:

- To educate the user
- To engender trust in the user
- To provide actionable insights to the user
- To provide insights regarding system design and debugging

Surprisingly, most XAI methods focus on serving the needs of AI engineers in system debugging, and this contributes significantly to the gap between explanations provided by AI models and those sought by end users.

In addition to serving the needs of the human-in-the-loop seeking the explanation (i.e. the explainee), an explanation should be presented in a simple and friendly manner. For example, existing datasets to train AI models for loan decisions contain features such as "external risk estimate, risk performance, trades with high utilization ratio", etc. Our study indicates that these factors are not helpful in justifying loan denials to applicants. Rather, users find specific factors concerning credit (no credit, limited credit, limited credit history), job (unstable job, unstable job history, no job), income (low income, no income) and debt (current debt, loan history) comprehensible and therefore acceptable.

Specifically, the objective of the study was two-fold.

- We wanted to curate a dataset that would aid in the creation of AI models which are capable of generating loan applicant friendly explanations such as those in (Srinivasan et al., 2018).
- We also wanted the dataset to include explanations that serve a purpose— explanations that educate the loan applicant about loan denial, and help them take appropriate actions to secure a future loan.

In the next section, we describe the process of curating the dataset.

## 2. How we built the dataset

Xnet is a unique and novel dataset that contains user-friendly explanations for loan denials. We highlight key features regarding data collection, content analysis and the explanation purposes served by the dataset below.

### 2.1. Snapshot

The dataset was curated from surveys conducted on Amazon Mechanical Turk (AMT) across multiple rounds in order to arrive at a reliable and valid set of responses. The use case was the same across all surveys: a loan applicant is seeking a $12000 loan from a bank to purchase a car. The bank uses an AI system to determine if the loan applicant is a good candidate. In this case, the applicant was denied the loan and is now seeking an explanation for how this decision was reached. The task of the AMT workers was to provide up to 10 explanations to the candidate the details of which is provided below.

**Data Collection**

- Data collected by means of survey on Amazon Mechanical Turk (AMT)
- AMT workers were provided with a loan application scenario and asked to imagine that they were loan applicants
- Workers provided textual descriptions highlighting reasons for loan denials

**Data Labeling**

The collected responses from AMT workers were edited for syntactic and semantic correctness by linguists. The responses were subsequently tagged with appropriate labels corresponding to the reason of loan denials. An analysis revealed that there were some high level reasons or broad categories of loan denials, and within each of those broad categories, several sub-categories could be identified.

- A total of 2432 responses were divided into broad categories (e.g., job) and subcategories (e.g., unstable job, no job) for loan denial.
- The analysis yielded a total of 16 major categories each of which contained 2-7 sub-categories. A total of 60 unique loan denial reasons were thereby identified.

**Expanding Explanations along Purpose Dimensions**

As outlined in the second objective, the next task in curating the dataset was to incorporate explanations that help educate the loan applicant about the denial, and to help them take appropriate action to secure a future loan. While the survey results offered a diversity of ways to express the explanations, we re-phrased those explanations, not only for the purposes of clarity, but also for the purposes of expressing sympathy and caring to the user. This, in turn, would facilitate the creation of a friendly channel of communication for educating and guiding them. Some illustrations are provided in Table 1.

- Explanations that educate the loan applicant
- Explanations that help the loan applicant take an appropriate action towards future loan approval

In the next section, we provide details concerning survey designs.

### 2.2. Survey Development

We varied the details that we provided to the survey respondents in order to obtain a range of useful explanations. In Survey 1, we provided specific details about the applicant. By providing these specific details, we hoped to provide a realistic portrait of the loan applicant that would inspire the survey responses. However, we quickly found that most of the responses focused solely on the information we had provided.

**Survey 1 (10 participants)**
*Unique survey feature:* Provided specific details about the loan applicant (e.g. $ 28000 annual income)
*Observations:* Responses were too biased towards the information provided about the applicant

To rectify this, we developed Survey 2 which replaced the specific details of the applicant with five broad categories of features that the AI might use in making its decision (e.g. credit score, annual income, etc.). However, we found again that the responses were heavily biased towards the features provided.

**Survey 2 (50 participants)**
*Unique survey feature:* Provided examples of features (broad categories such as credit history, etc.) that the AI might use in reaching its decision without listing specific numbers as in Survey 1.
*Observations:* Responses were biased by the features listed with no additional reasons for denial provided.

We therefore developed Survey 3 which eliminated both the specifics about the applicant and the sample features. As we hoped, the answers were unbiased and diverse across a number of categories.

**Survey 3 (50 participants)**
*Unique survey feature:* Provided no information about the applicant or the possible features that the AI model could use.
*Observations:* Responses were diverse with many unique explanations provided.

The final version of the survey, Survey 4, was identical to Survey 3 with the exception that we asked respondents to provide both good (i.e., useful) and poor (i.e., not useful) explanations. We added the poor explanations to learn about any additional explanation characteristics that might emerge from this category.

**Survey 4 (Final) (200 participants)**
*Unique survey feature:* Same as Survey 3 but respondents were asked to provide examples of both good and poor explanations
*Observations:* Responses were diverse across both good and poor explanations. This survey version was adopted for collecting the data.

We discovered 16 broad categories of explanation. This qualitative annotation could serve as the input for training the AI models.

In the next section, we summarize some of the findings of the study.

### 3. What we learned

The study revealed some interesting aspects.

- Existing datasets to train explainable AI models contain features that are mostly incomprehensible by end users. Furthermore, existing datasets seldom contained features that users considered acceptable in understanding a model's decision. For example, in understanding loan denials, users considered features such as assets, income, debt, etc. as acceptable. These features are hardly present in typical machine learning datasets to determine loan denials.

- Different people have different views of the same world. These subjective viewpoints, get baked into the dataset and, in turn, into the kind of explanations they seek. As long as these are valid reasons, these diverse viewpoints are essential to address the needs of various kinds of users. Incorporation of various viewpoints improves diversity of the dataset.

- As a consequence of the aforementioned point, there are often several explanations that appear only once in the dataset reflecting the viewpoints of certain individuals only. This kind of a data distribution poses challenges to machine learning algorithms due to sparsity of data points.

- Crowdsourcing data comes with its own challenges. The data has to be cleaned for semantic and syntactic correctness and labeled. This would involve another set of expert humans in the loop. In order to maintain consistency amongst the annotators, certain rules had to be established in editing and annotating. First, we decided to discard responses that were not well formed (for example, "the loan was denied because the applicant is poor" is not a well formed reason as it lacks the necessary details about what it means by being poor), or those that were biased (for example, "the loan was denied because the applicant is a woman"). For editing, certain explanation structures were adopted. Based on studies from cognitive science, we incorporated a few explanation structures (Keil, 2006) that are proven to be helpful to the explainee. For example, explanations were structured as if-then rules (for example, if the annual income is less than x, then a loan cannot be granted) and as causal patterns (for example, the loan is denied because the applicant's credit score is too low). The annotators then would re-frame each response based on those rules.

| Education | Action |
|---|---|
| The credit associated with this application is unfortunately not high to be considered elgible for this loan | Successful loan applications tend to have high credit associated with their account. So talk to your bank about finding ways to improve your credit |
| A look at the application shows that there are not enough assets listed to back up the loan | Please consider re-applying for a loan with assets of greater value than the loan requested. |

Table 1. Illustrations of explanations that educate and that help in suggesting appropriate actions

## 4. What we recommend

In order to bridge the gap between research and practice, and to accelerate the practical adoption of AI across a variety of domains, we suggest:

- *Adoption of a user-centered approach in generating explanations (rather than model-based approaches):*
  In other words, explanations need to serve a variety of users and therefore have to be designed in a way that no domain knowledge may be necessary. This, in turn, would mean that the reasons used in the explanation should be easily comprehensible by various stakeholders involved.

- *Understanding of the explanation context: who is seeking explanation, what is the purpose of the explanation, etc.*
  Explanations are meant to bridge the knowledge gap between the person providing the explanation (explainer), and the person seeking it (explainee). Furthermore, explanations should serve a purpose. This could be in educating the user, in engendering trust in the user, in helping with system debugging, to guide the user in taking appropriate actions, etc. Thus, it is important to design explanations keeping in mind the human-in-the-loop seeking it, and the purpose an explanation is trying to serve.

- *Creation of new datasets aligned along the lines of aforementioned points*
  As already described, features in existing machine learning datasets are not necessarily user friendly. Creating new datasets that are representative of user-friendly features to train new AI models is therefore beneficial. In creating such datasets, it is very much possible that one may encounter unfamiliar or uncommon explanations for a particular use case. It is important to include these uncommon responses as long as they are valid in order to capture diversity and to address the needs of various kinds of users.

- *Design of new algorithms to cater to the limitations of datasets*
  As a consequence of creating user friendly datasets, certain use-cases may not have ample training data. In such scenarios, new algorithms may have to be designed to handle limited training data.

## 5. Conclusions

Most explainable AI techniques are focused on designing algorithms that can explain an AI's decision. However, the data that is used to train these algorithms may contain features that are often incomprehensible to an end-user even with the best XAI algorithms. Thus, the problem of explainability has to be addressed starting right from the data creation step. In this paper, we studied this problem considering the use-case of explaining loan denials to end-users as opposed to AI engineers or domain experts. We described the process of curating such a dataset, and the insights gained from the study. We believe there is a pressing need for the adoption of a user-centric approach in both the creation of datasets and in the design of XAI algorithms. We hope our findings will serve a as guide to future researchers in the field.